\documentstyle[aps,prb,epsfig,float,twocolumn]{revtex}
\begin{document}
\twocolumn[\hsize\textwidth\columnwidth\hsize\csname
@twocolumnfalse\endcsname

\title{Spin decay and quantum parallelism}

\author{John Schliemann, Alexander V. Khaetskii, and Daniel Loss}

\address{Department of Physics and Astronomy, University of Basel,
CH-4056 Basel, Switzerland}

\date{\today}

\maketitle

\begin{abstract}
We study the time evolution of a single spin coupled
inhomogeneously to a spin environment. Such a system is realized
by a single electron spin bound in a semiconductor nanostructure
and interacting with surrounding nuclear spins. We find striking
dependencies on the type of the initial state of the nuclear spin
system. Simple product states show a profoundly different behavior
than randomly correlated states whose time evolution provides an
illustrative example of quantum parallelism and entanglement in a
decoherence phenomenon.
\end{abstract}
\vskip2pc]


\section{Introduction}

The interest in electron spin dynamics in semiconductor structures has 
remarkably increased in the recent years generating the emerging field
of spintronics \cite{Wolf01,Awschalom02}. This key word summarizes the efforts 
to use the spins of quantum objects rather than or in combination with their
charge for information processing, or, even more ambitious,
for quantum information processing. Meanwhile several proposals
for quantum information processing using (electron or nuclear) spins have been
put forward \cite{Loss98,Privman98,Kane98,Levy01,Ladd01}.

In quantum information processing the coherence of quantum bits is crucial.
This issue becomes particularly important in solid state systems where
the qubits are ususally assumed to be affected by much more and much stronger 
perturbating influences than in other experimental setups related
to quantum information processing in various other fields of physics such as
atomic physics, quantum optics, or NMR experiments. 
An important advantage however solid state systems have is that
they offer the possibility scalability once individual qubits and elementary
gate operations between them are estabilshed. Such a perspective is usually
not given in other quantum computation scenarios.

Motivated by these developments we investigate in this work the
time evolution of single spin $\vec S$ that is coupled
inhomogeneously to a non-interacting environment of other spins.
A natural realization of
such a system is given by the spin of a single
electron in a semiconductor quantum dot interacting with
surrounding nuclear spins via hyperfine coupling
\cite{Dyakonov84}. Alternatively one can think of an electron
bound to a phosphoros atom implantated into a silicon matrix
\cite{Kane98}, or of other combinations of impurities and host
materials such as Si in a GaAs or Ge matrix.
The time evolution of an
electron spin under such circumstances is of fundamental interest
in its own right and of particular relevance to the quantum
computing proposal of Ref. \cite{Loss98,Kane98}. In fact, very
recently a series of studies of electron spin dynamics related to
the present one has appeared
\cite{Khaetskii00,Erlingsson01,Mozyrsky01,Frasca01,Khaetskii02,Erlingsson02,deSousa02,Merkulov02,Saykin02}.
Here we build on  recent work in Ref.~\cite{Khaetskii02},
where the dynamics of an
electron spin due to the hyperfine interaction with nuclear spins in a
semiconductor quantum dot was investigated.
This scource for spin decay can be assumed to be 
the dominant one in a quantum dot geometry where other mechanisms induced
by spin-orbit interaction are believed to be suppressed, although this
issue has not been clarified entirely yet \cite{Mozyrsky01}.

Our approach here is based on numerical simulations of the full
quantum mechanical spin dynamics in sufficiently small systems.
These studies complement the earlier work in Ref. \cite{Khaetskii02},
where approximation-free analytical results where obtained for the case of a
fully polarized system of nuclei. For a more general initial condition
such as an unpolarized nuclear spin system low-order time-dependent
perturbation theory was employed which unfortunately suffers from
divergent terms in higher order.

In our numerical simulations we observe a decay of the electron spin
as measured in terms of the expectation values of its components.
We study this phenomenon as a function of the initial nuclear polarization,
and the type of the initial state of the nuclear spin system.
We compare the time evolutions of initial states where the nuclear
system is in a simple tensor product state with the situations where the 
initial nuclear state is randomly correlated. A major results is that the time
evolution depends very significantly on the type of initial state of the
spin environment. The time evolution of simple tensor
product states can be quite individual, while randomly correlated 
(and therefore highly entangled) states
show very reproducible dynamics that mimics the {\em average} over
the time evolutions of all possible tensor product states. This
observation is an example of {\em quantum parallelism} in
a decoherence phenomenon. 

The spin decay is adccompanied by the generation of quantum correlations 
between the electron spin and the nuclear spins, illustrating
a general concept of quantum information theory where the decoherece
of a quantum bit (here the decay of the electron spin) is viewed as
the result of the generation of entanglement (i.e. quantum correlations)
between the qubit and its environment. We quantify this entanglement using
well-established methods and concepts of quantum information theory.
By this we also hope that studies of this kind will faciliate 
fruitful interactions between the communities of solid state physics and 
quantum infromation.

Finally we compare the results
for the full quantum mechanical dynamics with simulations of a classical
spin model that arises as the classical limit of the underlying
Hamiltonian. This comparision shows that the spin decay
observed in the quantum system depends crucially on two properties
of the system: (i) the inhomogeneity 
of the hyperfine contact interaction induced by the spatial variation of the
electron wave function, and (ii) the quantum mechanical
nature of the dynamics allowing for nontrivial correlations 
(entanglement) between the electron spin and the nuclear spins.

This paper is organized as follows: In section \ref{model} we briefly
describe the details of our modeling and technical aspects of 
our numerical simulations.
In section \ref{simulations}
we report on our numerical results. We discuss the role of different
initial conditions for the nuclear spin system, and 
the connection between decoherence and the generation of 
entanglement observed in our simulations.  
We close with conclusions in section \ref{conclusions}.


\section{The model}
\label{model}

We consider single spin $\vec S$ which is coupled
inhomogeneously to a non-interacting environment of other spins,
\begin{equation}
{\cal H}=\vec S\cdot\sum_{i}A_{i}\vec I_{i} \label{defham}\,.
\end{equation}
The coupling is inhomogeneous since the constants $A_{i}$ vary
among the environment spins $\vec I_{i}$. For an electron spin
residing in semiconductor quantum dot the coupling constants
$A_{i}$ are given by $A_{i}=Av_{0}|\Psi(\vec r_{i})|^{2}$
where $A$ is an overall coupling parameter and $v_{0}$ the inverse 
density of nuclei in the material. $\Psi(\vec r_{i})$ is the electron
envelope wave function at location $\vec r_{i}$. This factor induces a
spatial dependence of the cupliong constants $A_{i}$ which is crucial for
the spin dynamics.
For simplicity we will consider in the following nuclear spin of length
$I=\frac{1}{2}$ in a spherical quantum dot
In our simulations a given number $N$ of nuclear spins
is contained in a sphere of radius $R=(3N/4\pi n_{0})^{1/3}$
where $n_{0}=1/v_{0}$ is the density of nuclei. The electron wave function is 
given by
\begin{equation}
|\Psi(\vec r)|^{2}=\left(\frac{1}{\pi (R/a)^{2}}\right)^{\frac{3}{2}}
e^{-\frac{r^{2}}{(R/a)^{2}}}
\end{equation}
where the parameter $a$ describes the confinement of the electron in the dot.
In the following we shall use $a=2$
and the material parameters of gallium arsenide with 
$n_{0}=45.55{\rm nm}^{-3}$. Therefore a typical quantum dot contains about
$N=10^{5}$ nuclei. For the alternative scenario of an
electron bound to a $^{31}{\rm P}$ in silicon the number $N$ of nuclear
spins effectively coupled to the electron spin is smaller. The
Bohr radius of the hydrogen-like electron state is about 3nm, and
with the lattice constant of silicon and the natural abundance of
$^{29}{\rm Si}$ this leads to values for $N$ of about a few hundred.

To mimic their sperical distribution also in systems
of smaller size used in our simulations
we choose the radial coordinate $r_{i}$ of the $i$-th nucleus according to
$r_{i}=[3(i-1/2)/(4\pi n_{0})]^{1/3}$ with
$i$ ranging from $1$ to $N$.

The Hamiltonian (\ref{defham}) does not include the direct dipolar interaction
between nuclear spins. This interaction is by orders of magnitude weaker 
than the scale $A$ of the hyperfine coupling which is of order
$10^{-5}{\rm eV}$ 
in GaAs \cite{Dyakonov84}. In this material the characteristic time
$T_{2N}$ for the nuclear spin decay due to dipolar interaction
is of order $10^{-4}{\rm s}$ while the time scales considered in this
work will be at least two orders of magnitude smaller. We also mention a
recent interesting numerical study by Dobrovitski {\sl et al.}
\cite{Dobrovitski01} on spin dynamics stressing the role of
entropy. There a central spin is coupled inhomogeneously to an
essentially non-interacting spin environment, where, differently
from the present study, an Ising-like coupling was used. To allow
for nontrivial dynamics the authors of Ref.~\cite{Dobrovitski01}
introduced a magnetic field perpendicular to the $z$-direction of
the Ising coupling.

The model (\ref{defham}) was specifically studied recently in
Ref.~\cite{Khaetskii02}, where approximation-free analytical
results were obtained for the case of a fully polarized system of
nuclei. For a more general initial conditions such as an
unpolarized nuclear spin system low-order time-dependent
perturbation theory was employed which unfortunately suffers from
divergent terms in higher order. In this work we choose a
different route and perform finite-size exact diagonalizations
from which we obtain the full time evolution.

Since our Hamiltonian conserves the total spin $\vec J=\vec
S+\sum_{i}\vec I_{i}$ it is convenient to work in a subspace of
given $J^{z}$ having a dimension of 
\begin{equation}
\left(
\begin{array}{c}
N+1 \\ \frac{N+1}{2}-J^{z}
\end{array}
\right)\,.
\end{equation}
To obtain the time evolution of the total spin system with the initial state
lying in a subspace with a given value $J^{z}$ we diagonalize the
Hamitonian within this subspace and compute the time evolution
of certain expectation values from the eigensystem data.
For initial states having non-zero overlap in several subspaces with
different $J^{z}$ the time evolutions obtained in the different spaces
have to be superimposed.
The fact that the full eigensystem of the Hamiltonian is required
in this procedure
is different from most other numerical investigations of spin systems 
where, for reason of the physical questions being investigated, 
it is sufficient to concentrate on the ground state and some low-lying 
excitations. In our case we need the full eigensystem and are therefore 
restricted to system sizes $N$ and values of $J^{z}$ yielding dimensions of
not more than a few thousands.  
However, as we will explain in more detail below,
our findings neither depend on the restriction to a certain value of
$J^{z}$ nor on the
specific coupling constants induced by the quantum dot geometry.
In fact, qualitatively the same results are obtained when working
in the full Hilbert space, or if the coupling parameters are
chosen at random from a uniform distribution. On the other hand it
is essential that the coupling is inhomogeneous leading to a time
evolution which is for all practical time scales aperiodic and in
this sense irreversible. The case of homogeneous coupling with all
$A_{i}$ being the same is readily solved analytically and
generates periodic dynamics with recurrence time $T=4\pi
N\hbar/A$.


\section{Results for the time evolution of spins}
\label{simulations}

In the time evolutions to be discussed below the initial state
will always be a simple direct product of the state of the nuclear
or environmental spin system and the electron spin with the latter
pointing downwards along the $z$-direction. Therefore the nuclear
spins and the electron spin are initially uncorrelated. 
For the nuclear spin system itself we consider two types of initial
conditions which give rise to significantly different time evolutions.

\subsection{Product states versus randomly correlated states}

We investigate two types of initial states for the nuclear spin system
which differ crucially in their corrrelation properties 
and, as a we shall see below, as a consequence of this
also in their time evolution:
(i) The nuclear spins are initially in a simple tensor
product state. If $J^{z}$ is fixed to a certain value such a state
consists of eigenstates of $I^{z}_{i}$ for each nuclear spin $i$.
If this restriction is not applied tensor product states
consisting of more general spin-coherent states are possible. (ii)
The nuclear spin state $|\chi_{N}\rangle$ is initially a  linear
superposition, $|\chi_{N}\rangle=\sum_{T}\alpha_{T}|T\rangle$,
where the sum goes over all tensor product states $|T\rangle$
consisting of eigenstates $I^{z}_{i}$, $i\in\{1,\dots,N\}$, and
is, for fixed $J^{z}$, restricted to the appropriate subspace. The
coefficients $\alpha_{T}$ in this  entangled pure state are
subject to a normalization condition and chosen either at random
or coherently (for example, can have the same phases).

\begin{figure}
\centerline{\includegraphics[width=8cm]{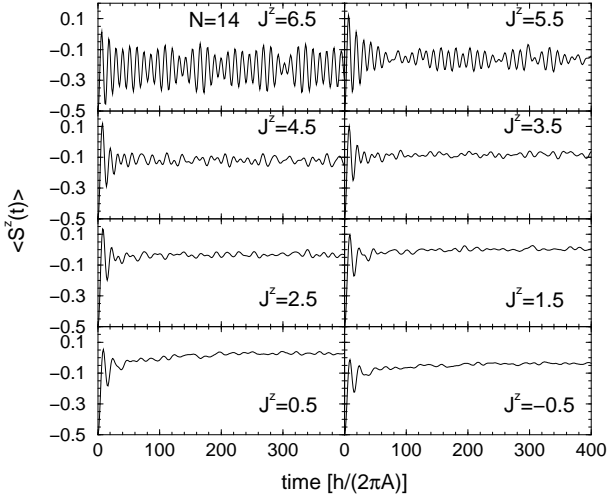}} \caption{The
time evolution of the electron spin in a system of $N=14$ nuclear
spins for different degrees of polarization of the randomly
correlated nuclear system and coupling constants  induced by the
quantum dot geometry. In the top left panel the nuclear spins are
fully polarized in the initial state with the electron spin
pointing opposite to them ($J^{z}=13/2$). In the following panels
the number of flipped nuclear spins in the initial state is
gradually increased. The case of an initially fully unpolarized
(but randomly correlated) nuclear system is reached in the bottom
right panel ($J^{z}=-1/2$). Here and in the following we take
spins to be dimensionless, i.e. measured in units of $\hbar$.
\label{fig1}}
\end{figure}

As we shall see shortly, a single tensor product state on the one hand
and a randomly correlated nuclear state on the other hand 
generate strikingly different time evolutions for the
electron spin. Fig.~\ref{fig1} shows numerical data for
time-evolved expectation value $\langle S^{z}(t)\rangle$ for an
initially randomly  correlated  system and different degrees of
its polarization (characterized by $J^{z}$). In all cases,
$\langle S^{z}(t)\rangle$ decreases in magnitude. With decreasing
polarization the decay becomes more pronounced, and the
oscillations accompanying this process get suppressed. Note that
it is the decay of the envelope in these graphs but not the 
fast oscillations itself that signals the decay of the spin.
The distance between two neighboring maxima of the oscillations
can depend slightly on the initial state and the coupling constants
in the Hamiltonian. However, a good estimate for this effective period is
usually given by $T=4\pi\hbar/A$ since $A/2$ is an estimate
(neglecting quantum fluctuations) for the width of the spectrum,
i.e. the difference between the largest and the smallest eigenvalue 
of the Hamiltonian.

\begin{figure}
\centerline{\includegraphics[width=8cm]{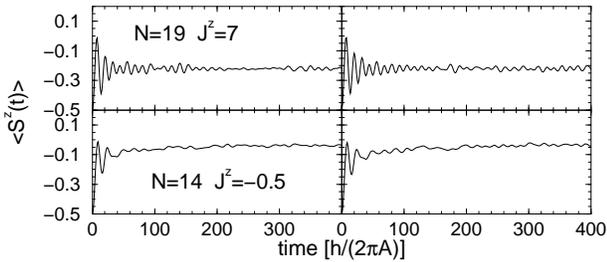}} \caption{Upper
panels: $\langle S^{z}(t)\rangle$ for a system of size $N=19$
being initially in a  randomly correlated nuclear spin state in
the subspace with $J^{z}=7$. The two panels represent two
different randomly chosen initial conditions. Lower panels:
Analogous data for $N=14$ and a completely unpolarized nuclear
spin system ($J^{z}=-1/2$). In both cases the simulation data does
practically not depend on the initial condition. \label{fig2}}
\end{figure}

When the nuclear spin system is initially in a randomly correlated
state the time evolution of $\langle S^{z}(t)\rangle$ is very
reproducible in the sense that it depends only very weakly on the
particular representation of the initial random state. This is
illustrated in Fig.~\ref{fig2} where the results of different
initial random configurations are compared for two different
system sizes and degrees of polarization.

\begin{figure}
\centerline{\includegraphics[width=8cm]{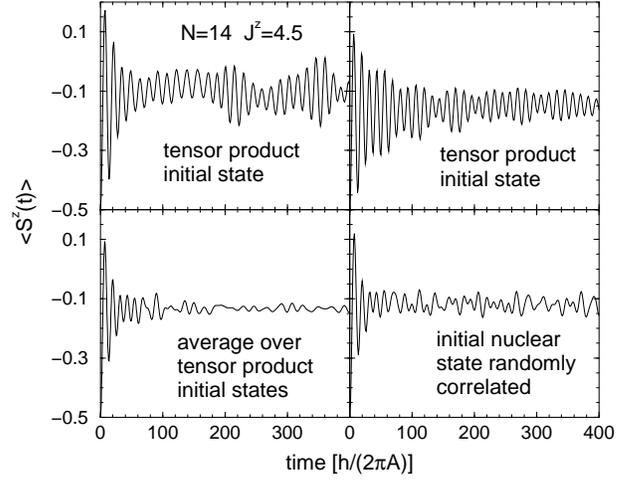}} \caption{Upper
panels: time evolution of the electron spin $\langle
S^{z}(t)\rangle$ for a system with 14 nuclear spins being
initially in an uncorrelated tensor product state in the subspace
 $J^{z}=9/2$. The oscillation period
  and the time scale of the decay are consistent with the period $ T=4\pi
\hbar/A$ and the  scale $\hbar N/A$ identified in
\protect\cite{Khaetskii02}.
 Lower left panel: data
of the same type as above but averaged over all possible
uncorrelated initial states with $J^{z}=9/2$. Here again, the time
scale of the decay is consistent with the scale $\hbar \sqrt{N}/A$
identified in \protect\cite{Khaetskii02}. Lower right panel:
$\langle S^{z}(t)\rangle$ for the same system being initially in a
randomly chosen correlated state. \label{fig3}}
\end{figure}

This behavior of randomly correlated initial states sharply
contrasts with the time evolution of simple tensor product nuclear
spin state. The upper two panels of Fig.~\ref{fig3} show the time
evolution of the electron spin for two initial tensor product
states. In the lower right panel we compare this data with the
time evolution of a representative of the randomly correlated
initial condition. In the former case the time evolution depends
significantly on the concrete initial tensor product state, and
the decay of the electron spin occurs typically clearly more
slowly than in the case of an initially randomly correlated
nuclear spin system.

In the lower left panel of Fig.~\ref{fig3} we show the time
evolution of the electron spin averaged over all nuclear tensor
product states \cite{note1}. Comparing the two lower panels one
sees that this data is very close to the time evolution of a
randomly correlated state. This observation is also made for other
system sizes and degrees of polarization and constitutes an
example of {\em quantum parallelism} \cite{Steane98}: 
The time evolution of each
initially uncorrelated (and therefore classical-like) nuclear
state is present in the evolution of a linear superposition of all
such states. In other words, the time evolutions of all
uncorrelated classical-like states are perfomed in parallel
in the time evolution of the randomly correlated state.
An experimental consequence of this observation is
that if the electron spin dynamics would be detected on an array
of independent quantum dots one could not distinguish whether the
nuclear spin system in each dot was initially randomly correlated
or in an uncorrelated tensor product state. In other words, the
spin dynamics of a  randomly correlated pure state of the nuclear
system in a single dot cannot be distinguished from a mixed state
of an ensemble of dots.

\begin{figure}
\centerline{\includegraphics[width=8cm]{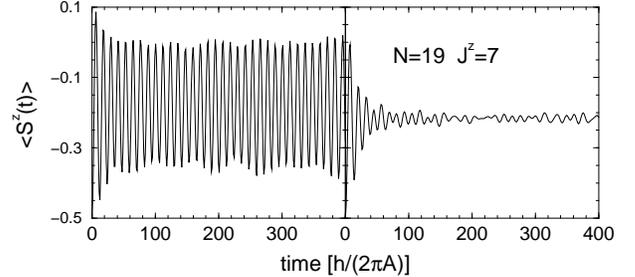}} \caption{Time
evolution of $\langle S^{z}(t)\rangle$ for two types of initially
randomly correlated nuclear spin states. In the left panel the
amplitudes $\alpha_{T}$ are restricted to have non-negative real
and imaginary part, while in the right panel they have all the
same modulus but completely random phases. \label{fig4}}
\end{figure}

The observation that the time evolution of a randomly correlated
state quite closely mimics the average over all tensor product
initial conditions relies on the cancellation of off-diagonal
terms $\alpha_{T}^{\ast}\alpha_{T'}\langle \Downarrow,T|\vec
S(t)|\Downarrow,T'\rangle$, $T\neq T'$, due to the randomness in
the phases of the coefficients $\alpha_{T}$. In this sense our
system has a self-averaging property. This can be checked
explicitly by reducing this randomness. The left panel of
Fig.~\ref{fig4} shows the time evolution of a randomly correlated
state where the amplitudes $\alpha_{T}$ are restricted to have a
non-negative real and imaginary part. This time evolution turns
out to be similarly reproducible as before, i.e. it does not
depend on the concrete realization of the initial random state,
but it is clearly different from the former case since the
cancellation of off-diagonal contributions is inhibited
\cite{note1}.
For comparison we show in the right panel data where
the amplitudes in the initial nuclear spin state have a random
phase but are restricted to have the same modulus. Here the proper
averaging process takes place again.
The results described so far were
obtained in certain subspaces of $J^{z}$ and for the form of
coupling constants $A_{i}$ as induced by the quantum dot geometry.
However, our findings do not depend on these choices. We have also
performed simulations were the initial state has overlap in the
full Hilbert space. For a randomly correlated initial nuclear spin
state the only difference is that now also transverse components
$\langle S^{x}(t)\rangle$, $\langle S^{y}(t)\rangle$ of the
electron spin evolve. However, these are tiny in magnitude and
oscillate around zero. For an initial tensor product states these
transverse components can become sizable, and the time evolution
again strongly depends on the concrete initial tensor product
state. Moreover, as mentioned earlier, the exact form of the
coupling constants is also not crucial as long as they are
sufficiently inhomogeneous. For instance, we obtain qualitatively
the same results if we choose the coupling parameters randomly
from a uniform distribution.

We also note that coupling a magnetic field to the electron spin
has only a quantitative influence on our results. Here again the
time dependence of tensor product initial nuclear state is very
individual, while a randomly correlated states gives very
reproducible results that mimic closely the average over tensor
product states.

\subsection{Decoherence and the generation of entanglement}

In circumstances of quantum information processing the decay of a
qubit is ususally viewed as some 'decoherence' process due to the
environment attacking the quantum information. As seen above, the
spin decay is generically slower if the spin environment is
initially in a uncorrelated state. This finding suggests that it
is advantageous for protecting quantum information to disentangle
the environment that unavoidably interacts with the qubit system.

A 'decoherence' process of the above kind can be viewed as the
generation of entanglement between a qubit and its environment.
The system investigated here provides an illustrative example for
this statement. The entanglement in the total state
$|\Psi(t)\rangle$ between the central electron spin and its
environment can be measured by the von-Neumann entropy of the
partial density matrix where either the electron or the
environment has been traced out from the pure-state density matrix
$|\Psi(t)\rangle\langle\Psi(t)|$ \cite{Bennett96}. Tracing out the
nuclear system we have
\begin{equation}
\rho_{el}(t)=\left(
\begin{array}{cc}
\frac{1}{2}+\langle S^{z}(t)\rangle & \langle S^{+}(t)\rangle\\
\langle S^{-}(t)\rangle & \frac{1}{2}-\langle S^{z}(t)\rangle
\end{array}
\right) \, .
\end{equation}
This matrix has eigenvalues $\lambda_{\pm}=1/2\pm|\langle\vec
S(t)\rangle|$, and the measure of entanglement reads
$E(|\Psi(t)\rangle)=
-\lambda_{+}\log\lambda_{+}-\lambda_{-}\log\lambda_{-}$. Thus, the
formation of expectation values $|\langle\vec S(t)\rangle|\neq
1/2$ (or, in the case of fixed $J^{z}$, just $|\langle
S^{z}(t)\rangle|\neq 1/2$), is a manifestation of the entanglement
between the electron spin and the nuclear spin system. The maximum
entanglement, $E=\log 2$, is achieved if the electron spin has
decayed completely as measured by the expectation values of its
components, $\langle\vec S(t)\rangle=0$. The generation of quantum
entanglement between the electron spin and the nuclear spin system
signalled by a reduced value of $\langle\vec S(t)\rangle$ is a
main and crucial difference between the quantum system studied
here and its classical 'counterpart' described by a system of
Landau-Lifshitz equations. These equations can be obtained
from the Heisenberg equations of motion for the quantum system,
$\partial\vec S/\partial t=i[{\cal H},\vec S]/\hbar$,
$\partial\vec I_{i}/\partial t=i[{\cal H},\vec I_{i}]/\hbar$,
by performing expectation values of both hand sides
within spin-coherent states and assuming that the expectation values
of all operator products factorizes to products of expectation values.
This procedure becomes exact in the classical limit
\cite{Schliemann98}. The resulting equations do not contain 
operators any more but just describe the dynamics of three-component
vectors (classical spins) of fixed length.  We have performed simulations 
of such a classical spin system
by solving the Landau-Lifshitz equation via the fourth-order Runge-Kutta
scheme. As a result, the central classical spin performes an irregular
chaotic motion which does not show any similarity to the results
for the quantum spin-$\frac{1}{2}$ case.
In particular all qualitative features of 
quantum effects such as the generation of entanglement (signalled by
a decay of spins as measured by their expectation values) are not
present in such a time evolution.
Therefore the Landau-Lifshitz equation 
provides only a rather poor description of the
underlying quantum system.

Let us now briefly discuss how the different initial conditions
can be prepared experimentally. A tensor product initial state can
be produced by applying a magnetic field and having the underlying
crystal lattice at a temperature high enough such that
spin-lattice relaxation processes to the nuclear spins are
efficient. These interaction with the phonon environment will
effectively perform projection-type measurements on each spin and
force the system to be in a state close to a tensor product of
nuclear states pointing in each of the two direction along the
field axis. Another possibilty is the use of all-optical NMR
techniques as described in Ref.~\cite{Salis01}. A randomly correlated
nuclear state on the other hand can be achieved by cooling down
the lattice to temperatures where phonon processes are suppressed.
Then the highly anisotropic and long-ranged dipolar interaction
will produce a sufficiently 'disordered' state with a highly
irregular pattern of amplitudes when expressed in the tensor
product basis, as we have confirmed by explicit simulations of a
system of 8 nuclear spin placed on the edges of a cube. The highly
correlated (or entangled) character of these states can be
detected by following the individual nuclear spins in terms of
their expectation values $|\langle\vec I_{i}(t)\rangle|$. This
quantity decays from its initial value of $1/2$ (in a tensor
product initial state) on a time scale determined by the dipolar
interaction \cite{note2} to values typically close to zero.
According to the entanglement measure $E$ discussed above this
indicates strong entanglement between each nuclear spin $\vec
I_{i}$ and its environment of all other nuclear spins. In both
cases the initial state of the full system can be prepared by
injecting the electron to the quantum dot from an external lead,
or the electron state can be prepared by cooling in a magnetic
field and ESR techniques.

We finally consider the nuclear spin correlator $C(t)=\langle
I^{z}(t)I^{z}(0)\rangle$, $\vec I=\sum_{i}\vec I_{i}$, which can
be measured directly by local NMR like measurements such as
magnetic resonance force microscopy \cite{Suter02}. In a subspace
of given $J^{z}$ and the electron spin pointing downwards
initially this quantity reads $C(t)=(J^{z}-\langle
S^{z}(t)\rangle)(J^{z}+1/2)$. A realistic initial state will have
its dominant weight in a series of subspaces with neighboring
$J^{z}$ centered around some value. Then the time evolution of
$\langle S^{z}(t)\rangle$ is very similar in these subspaces, and
the dynamics of the total nuclear spin can be mapped out by
measuring the electron spin and vice versa.

\section{Conclusions}
\label{conclusions}

In summary we have studied the dynamics of a single spin coupled
inhomogeneously to a spin environment. As a main result the time
evolution depends sensitively on the type of initial state of the
spin environment. While the time evolution of simple tensor
product states can be quite individual, randomly correlated states
show very reproducible dynamics that mimics the {\em average} over
the time evolutions of all possible tensor product states. This
observation constitutes an example of {\em quantum parallelism} in
a decoherence phenomenon. This effects is clearly seen for all
finite system sizes studied here and can therefore be expected to
be also present in realistic quantum dot systems containing 
about $N=10^{5}$ nuclei, and also in the thermodynamic limit 
$N\to\infty$.  

The decay of the single spin in terms of
its expectation values is due to the formation of {\em
entanglement} between this spin and its environment. Since this
decay is generally slower if the spin environment is initially in
a simple tensor product state (i.e. no entanglement among the
environmental spins) our results suggest that it is advantagous
for protecting quantum information to {\em disentangle} the
environment. We expect this result to be of quite general nature, i.e.
it should also be valid for other systems consisting of some
central quantum object coupled to a bath of other quantum degrees of
freedom.

\acknowledgements{
This work has been supported by the Swiss NSF, NCCR Nanoscience,
DARPA, and ARO.}


%


\begin{references}
\bibitem{note1}
We note that averaging the operator $\vec S(t)$ over all tensor
product states for the nuclear system with the electron spin
pointing downwards is not the same as performing a trace over the
appropriate $J^{z}$ subspace since states with the opposite
direction of the electron spin are left out.
\bibitem{note2}
 The fact that the randomly correlated state has self-averaging
property means that even for a single dot with such a nuclear
state the time evolution of the electron spin can be described by
a statistical operator (mixed state) corresponding to an infinite
temperature of the nuclear system. We stress, however, that this
state is pure and possesses some additional properties (due to its
entanglement) which cannot be described by a  statistical operator
corresponding to some temperature.
\bibitem{note2}
This time scale is essentially given by the $T_{2n}$ for a bulk
nuclear system.
\end{references}
\end{document}